# The Enemy Within: The Emerging Threats to Healthcare from Malicious Mobile Devices

*Shams Zawoad and Ragib Hasan*
*University of Alabama at Birmingham*
*{zawoad, ragib}@cis.uab.edu*

## Abstract

With the proliferation of wireless networks, mobile devices and medical devices are increasingly being equipped with wireless interfaces, such as Bluetooth and WiFi to allow easy access to and control of the medical devices. Unfortunately, the very presence and usage of such interfaces also expose the medical devices to novel attacks from malicious parties. The emerging threat from malicious mobile devices is significant and severe, since attackers can steal confidential data from a patient's medical device. Also, attackers can compromise the medical device and either feed doctors bad data from it or issue potentially fatal commands to the device, which may even result in the death of the patient. As the mobile devices are often at close proximity to the patient (either in the hospital or home settings), attacks from such devices are hard to prevent. In this paper, we present a systematic analysis of this new threat from mobile devices on medical devices and healthcare infrastructure. We also perform a thorough security analysis of a major hospital and uncover potential vulnerabilities. Finally, we propose a set of potential solutions and defenses against such attacks.

## 1 Introduction

In recent years, medical devices are increasingly being equipped with wireless communication interfaces for better monitoring and controlling of patient's health. Infusion pumps, patient monitors, electrocardiograms, portable CAT machines, pulse oximeter sensors, heart rate monitors, pacemakers, breathing sensors, cardiac defibrillators, activity sensors, and insulin pumps are some examples of medical devices which have Bluetooth or WiFi access. Some of these devices are wearable, while some of them are implantable. According to ABI research, the market of wearable wireless devices will grow up to US$400 million by 2014 [2]. According to a study by the Freedonia Group, demand for Implantable medical devices (IMD) in the United States will increase 7.7% annually and will grow to a US$52 billion market by 2015 [16]. Besides that, smart mobile devices, such as, smart phones or tablets are also getting ubiquitous very rapidly which have the capability to communicate with wireless medical devices through the Bluetooth or WiFi channel. Leijdekkers et al. showed that it is possible to create a personal heart monitoring and rehabilitation system using smart phones [25]. This capability facilitates the growth of healthcare applications on smart mobile devices. Another ABI research report says that smart phone health applications will exceed US$400 million annually by 2016 [3]. The proliferation of wireless enabled mobile devices and medical devices has blessed our everyday life as well as the healthcare facility.

However, adversaries can misuse these Bluetooth and WiFi channels on mobile devices to launch novel attacks on medical devices. An attacker can get access to the patient's electronic medical record (EMR) by sniffing on the communication channel. She can also launch active attacks – for example, remotely issue a command for lethal doses on an insulin pump, or send stop command to an implantable pacemaker. The presence of mobile devices *inside* the security perimeter of the medical network makes them an attractive attack vector. Rather than being in close proximity to the patient, an attacker can launch an attack by proxy through a malware-infected mobile device belonging to the patient, a physician, or visitors. The attacker can also trigger attacks at specific time to have the most impact on the patient or her medical devices. In many cases, people who are using wearable or implantable medical devices use their smart phones to monitor their health conditions. As people tend to keep their smartphones with them almost all the time [30], there is high chance that a mobile malware can monitor the patient's condition 24x7. Through the mobile malware, an attacker can also target a specific group of people or a specific medical facility. Attacks from mo-bile devices, therefore, are significant and carry a greater risk to patient safety.

Unfortunately, it is hard to block the malware from using the network as many of these devices are already authenticated and authorized to use the required network. By the time we identify that the anomaly in medical device is due to a malware-infected mobile device, it may be too late to prevent the damage to patient health. Moreover, the resource limitation in both mobile and medical devices is the most pivotal hurdle in making these devices secured. Due to the computational and battery limitations, we can neither use traditional anti-malware software in medical and mobile devices, nor use strong encryption scheme for communication. Researchers have previously explored some attacks on medical devices and proposed network security solutions. But, attacks from mobile malware have not been fully explored yet and the schemes to secure traditional network communication are not always applicable to low-powered medical devices. And most importantly, the existing network security schemes do not consider insider attacks from compromised devices, where the enemy is already inside the secured region. In this paper, we explore a set of active and passive attacks on medical devices from an entirely novel medium – smart mobile devices. We identify the attack surface, perform a real-life case study, and explore mitigation strategies against these threats.

**Contributions:** The contributions of this paper are as follows:

1. We provide a systematic analysis of the threats from mobile devices to medical devices and infrastructure to identify vulnerabilities in current medical device usage models.

2. We perform a case study in a large hospital environment and identify vulnerabilities.

3. We explore mitigation strategies for securing medical devices against such attacks.

The rest of the paper is organized as follows. Section 2 provides the background and motivation of our work and presents our threat model. Section 3 describes the possible threats. In section 4, we state one case study and in section 5, we provide some solutions to prevent these attacks. Section 6 reviews prior works, relevant to our research and we discuss future work and conclusion in Section 7.

## 2 Motivation and Threat Model

In this section, we present our motivation for exploring possible threats from smart mobile devices to new generation Bluetooth or WiFi-enabled medical devices. We also discuss the threat model.

### 2.1 Motivation

While mobile malware and attacks on medical devices are not new phenomena, the combination of these two things can be devastating. Availability of WiFi and Bluetooth has made mobile devices more powerful than they ever were. However, at the same time, this has also increased the risks of new ways of attacks. Botnet command-and-control networks have used Bluetooth as a communication medium [36]. Android phones can also be used to exploit Bluetooth interface to hack into a car's Electronic Control Unit [12]. A lot of other attacks from smart phones are stated by Guo et al. [18]. Medical devices are also not safe from various kinds of attacks. It is possible to launch a software radio attack on implantable cardiac defibrillators and pacemakers [20]. Devices, such as, insulin pumps and glucose meters have also been successfully hacked [10, 27].

As both smart mobile devices and medical devices are now equipped with Bluetooth and WiFi interfaces, attackers can launch attacks on medical devices from compromised mobile devices. The consequence of such attacks on medical devices ranges from threats to the patients reputation to even her life. Sniffing EMR from the communication channels can be highly attractive to attackers because of the business value of medical records. Some attacks, such as providing wrong information to display devices or sending malicious command to some Implantable Medical Devices (IMD) or wearable devices, can actually place the patient's life in jeopardy. Murderers can use this as their killing tool as it is quite impossible to investigate [17]. Moreover, mobile malware based attack can open the opportunity of launching localized attacks. For example, adversary can target the topmost employees of an organization and if they wear any WiFi Real Time Location Service (RTLS) enabled device, she can monitor their location continuously. Therefore, it is important to systematically analyze the threats to medical devices from smart mobile devices and develop mitigation strategies.

### 2.2 Threat Model

To illustrate our threat model, we rely on the *Bring Your Own Device* (BYOD) model [7]. This model has attracted the attention of corporate security researchers exploring the enterprise security problem space [6]. Mod-ern office managers let their employers bring their own smart devices, such as, laptop, smart phone, tablet, etc.. The downside is that, an infected tablet or smart phone belonging to an executive can easily enter the system and the malware may thus be running inside the security perimeter of the corporation. In a healthcare setting, this will involve a mobile device belonging to a doctor, patient or a visitor who is present inside a healthcare facility or a person's house. Even if the patient is using strong security for her computing infrastructure, the infected device will likely be in the vicinity of the medical device. We assume that these mobile devices have already been infected by malware. This is possible when users download an untrusted application from the application store [15]. It is also possible to embed a malware in a trusted application or push a malware through an update patch of an existing application.

Pairing mechanism in Bluetooth channel offers most of the security of this channel. We assume that the malware affected devices are already paired with the Bluetooth-enabled medical device. This assumption is fairly common in malware research since people usually have many of their own mobile devices paired for their convenience. For example, while hacking a car's control unit, Checkoway et al. used an Android phone which was already paired with the car [12]. In the medical environment, this is more likely, where mobile devices belonging to healthcare personnel and/or patients are usually paired with the medical devices for their own interest (for example, a mobile bloodpressure app paired with a wearable blood pressure monitor). This assumption is sufficient to launch a Denial of Service (DoS) attack. We also assume that the attacker is capable of reverse engineering the application protocol of the Bluetooth-enabled medical devices. This can make her capable of learning patient data as well as sending incorrect signal to medical devices.

To attack through the WiFi channel, we assume that the malware infected mobile devices are already connected to the access point and the attacker can reverse engineer the application protocol. These assumptions are sufficient for DoS attack and to interfere the communication between mobile devices and WiFi-enabled medical devices.

## 3 Threat Taxonomy

We classify the threats into three main categories. Some attacks violate the privacy and confidentiality of patient's EMR, some break the integrity of communication between sensor device and display device or control device, while other attacks can affect the availability of the communication channel and the sensor device.

Figure 1 illustrates some possible attacks on healthcare devices from smart mobile devices

### 3.1 Privacy and Confidentiality

Nowadays, wireless medical devices are used to monitor patients' heart rate, activity, blood sugar, blood pressure and many other physical conditions. Physicians and patients can view these data from their smart mobile devices. Some implantable devices are helping doctors to monitor the patients condition remotely. Patients smart devices can upload the EMR to a centralized system, from where the physician get real-time information about the patients current condition remotely. According to the Health Insurance Portability and Accountability Act (HIPAA), these electronic medical data are private and confidential to the patient. People are also concerned about the privacy of their clinical data [34]. According to National Committee on Vital and Health Statistics (NCVHS), privacy means control over the acquisition, uses, or disclosures of personally identifiable health data, and confidentiality means obligations of those who receive information to respect the privacy interests of those to whom the data relate [13]. Using smart mobile devices, an adversary can exploit the available Bluetooth or WiFi channel to gain unauthorized access to the EMR and thus can violate privacy and confidentiality. Below are some possible attacks on privacy and confidentiality, we have identified:

**Breach of Medical Data:** Adversary can steal someone's medical data by Man-in-the-middle (MITM) attack on WiFi or Bluetooth channel. MITM in both Bluetooth and WiFi channel is well explored [19, 21]. If the sensor device and the control/display do not use any encryption or use weak encryption in their communications, then it is possible for an adversary to eavesdrop and sniff medical data. Stealing secret medical data and publishing it publicly can be a real threat to celebrities who do not want to disclose their physical problems.

We also note that the adversary need not be present near the patient by herself. She can create a malware which will collect data and upload it to her server. If the mobile devices of the patient, physician, or patient's visitors have this malware, then she can get the data via the malware. With numerous malware infected smart devices, she can compile a large electronic medical record database.

**Location Tracking:** WiFi RTLS system is also getting popular day by day. This location tracking technology has a tremendous effect in improving patient care, reducing costs, improving physical security and management of inventory. However, a malware infected phone can gather the location of a patient if she uses a location-tagged infusion pump. In the same way, the location of medical personnel can be exposed if they wear WiFi badges and their smart phones are infected by a malware. As people always tend to keep their smart phone with them, it is possible to track a persons location 24x7 in this way.

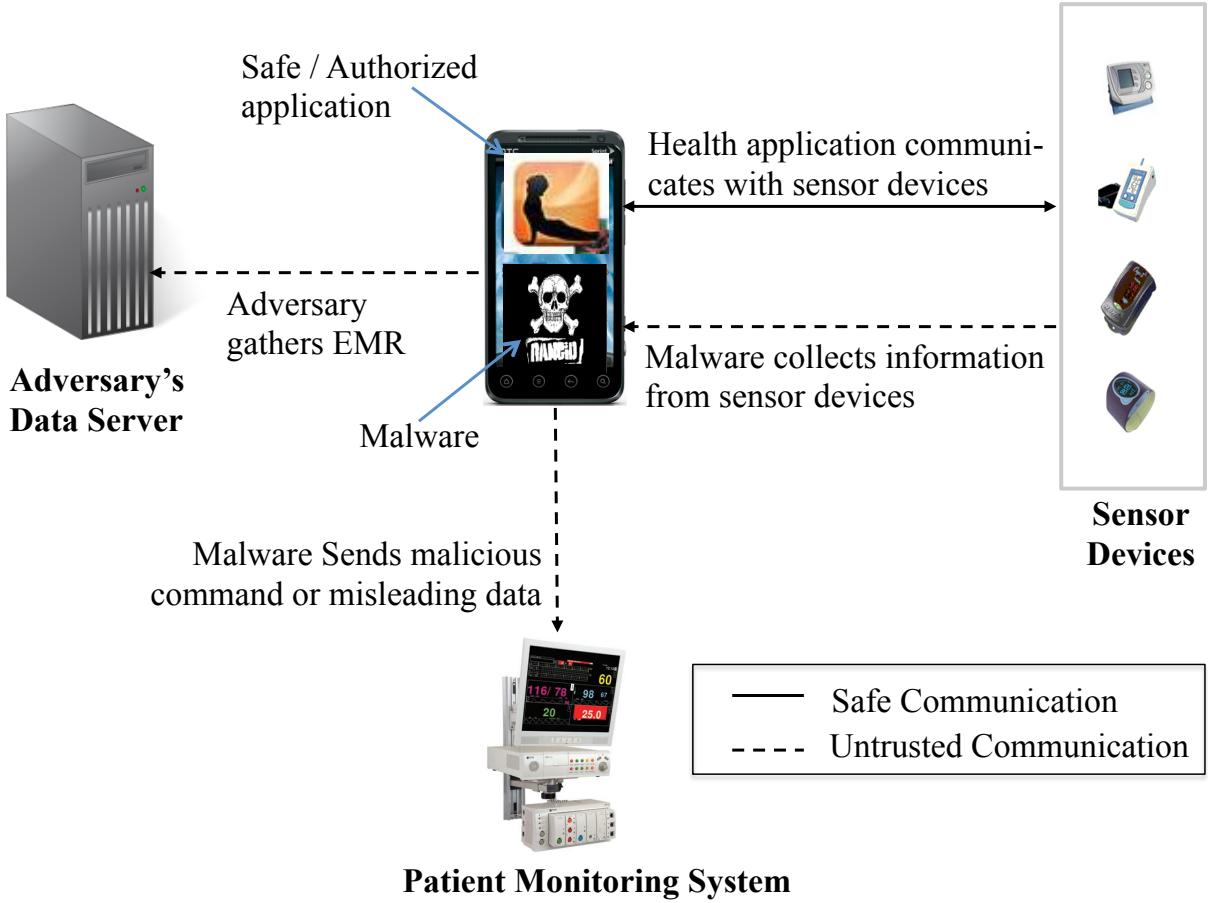

**Figure 1:** Attacks on Medical Devices from Smart Mobile Devices.

**Exposing Confidential Health Information:** Someone wearing a pacemaker or insulin pump may not want to expose that she has heart disease or diabetics to other people. But, if the devices have Bluetooth or WiFi access, it is possible for some adversary to know about the presence of these devices and thus know about the health condition of the person. Every WiFi or Bluetooth enabled device has a unique 48 bit MAC address which is visible to packet sniffers. Among the 48 bits, the first 24 bits are reserved for device manufacturer; it is called Organizationally Unique Identifier (OUI) [23]. And from this identifier, it is possible to identify device type. For example, from the MAC address we can distinguish between HTC EVO and HTC Hero, two Android phones from HTC [40]. Hence, it is possible to build a similar type classifier for medical devices. Then, an adversary will be able to know about the presence of any wearable and implantable device in a person's body or the surrounding environment.

### 3.2 Integrity

The communication channel between sensor device and control devices must be secured enough to provide data integrity. Otherwise, an adversary can come in the middle and send incorrect signal to medical/mobile devices. Below we describe some attacks which violate the integrity of the communication channel:

**Misleading Information:** By reverse engineering the application protocol, an adversary can use a smart phone to communicate with sensor or display device and feed incorrect data. So we can get incorrect blood pressure or glucose reading due to a malware installed on our smart phone. Sufficiently plausible incorrect data can lead doctors to take improper steps.

**Malicious Command:** A more dangerous attack will be sending incorrect command to a sensor device. A malware from patient's or a visitor's phone can send wrong command to a wearable or implantable device. It can be a life threatening attack, if the adversary is able to send malicious command to a implantable defibrillator, pacemaker or an infusion pump. For example, a malware installed in a smart phone can send a command of lethal dose to the patient's insulin pump.

## 3.3 Availability

Availability is a critical issue for medical devices. For some devices, such as, implantable pacemaker or cardioverter defibrillators, a short unavailability can be fatal to the patient. Here are some possible attacks on the availability of the wireless medical devices:

**Denial of Service:** The WiFi channel is based on IEEE802.11 [22] standard. It is well known that this standard is prone to DoS attack [8, 29] and interception [11]. Noubir et al. showed that the absence of error correction scheme has made the IEEE802.11 and Bluetooth vulnerable to DoS attack [29] .

Through our experiments, we have confirmed that it is possible to launch a DoS attack on the WiFi and Bluetooth channel by Android phone. If we know the external IP of a WiFi enabled device, then we can send a burst of packet from an Android phone to that device. Which in turn, can make that device unavailable for a while. From that observation, we can say that it is possible to make a medical device unavailable by using a smart mobile device.

**Battery exhaustion:** Another potential attack on the availability of medic devices is to drain out the battery of a device by keeping the communication channel busy continuously. Battery exhaustion or resource depletion attack on mobile devices through Bluetooth and WiFi channel is a real threat [28]. As new generation medical devices and mobile devices have similar communication channels, we predict that such attack on medical devices will be more common.

## 3.4 Localized Targeted Attack

We also look into the possibility of localized targeted attack from mobile devices. Such attacks can fall into any of the three above types. Someone can target a hospital to bring down their reputation by triggering the malware whenever doctors or patients come to that particular hospital. Attackers can also simultaneously trigger many attacks at a given hospital or even different areas of a city, causing many people requiring immediate medical attention at the same time and resulting in delays in emergency response.

## 4 Case Study

To explore the current security state of typical medical devices used in hospitals, we analyzed the system of a large university healthcare facility. Our tests found several vulnerabilities in the ongoing usage model for medical devices.

**Current Security Measures:** We found two security measures taken by the medical. Firstly, to avoid any intrusion in WiFi network of medical devices, the hospital maintains a separate WiFi network for medical devices and users. Patient or medical personnel can not use this WiFi network, dedicated for medical devices. And secondly, to ensure the data security, for some medical devices, Secured Socket Layer (SSL) is used to communicate with the central patient monitoring or EMR management system.

**Vulnerabilities and Possible Attack Scenarios:** Separate WiFi network for medical devices and smart mobile devices can secure the system against integrity violation, but still it is possible to attack on privacy and availability. Also, for personal health monitoring devices used at a home or workplace setting outside healthcare facilities, this precaution will be invalid since users need to connect their mobile devices with the medical devices via their home or work networks to observe their health condition. Even in the hospital, we were able to sniff the packets without connecting to the WiFi network. From the broadcast packets, we got the MAC and IP address of the surrounding devices and the wireless access points. By masquerading the IP address, it is possible to launch a DoS attack. The MAC address can also be used to identify device types and thereby infer the patient's confidential illness.

We identified insecurity in general purpose pumping machines, which communicate with the central patient monitoring system over the SSL through a gateway. They convey patients health condition to the central monitoring system, which returns appropriate dose of medicine to the pumping machines. Though it is not possible to intrude into a SSL communication, it is still prone to DoS attacks. Also, the gateway IP address was hard-coded in the device and was detected by our packet sniffer. Jamming this IP is possible which will make the communication channel unavailable. We cannot prevent this without updating the firmware as the gateway IP address is not configurable.

We also explored the vulnerabilities of medical devices, which use WiFi but do not communicate over SSL. One such machine is vital sign (body temperature, pulse rate or heart rate, blood pressure, and respiratory rate) monitor system. It communicates with the central EMR system over WiFi, without using SSL, even though it carries sensitive health information and Nurse's login credentials. Nurses need to login to this device to send information to the EMR system. As we were able to get the Address Resolution Protocol (ARP) packets from the channel, we could potentially launch a MITM attack by ARP spoofing to get all the data packets sent for the EMR system [14]. That device supports both Wired Equivalent Privacy (WEP) and WiFi Protected Access 2 (WPA2) encryption scheme. We can compromise the WEP encryption and get the actual EMR from this channel [39]. Using some tools, such as Aircrack [1], one can also learn about the patients' EMR. Nurse's login credentials can be exposed in the same way, which will be extremely dangerous for the whole EMR management system, since the adversary can gather anyone's confidential medical data by posing as a nurse.

Our brief analysis confirms our original arguments regarding the safety of WiFi and Bluetooth enabled medical devices. The vulnerabilities we exposed show that such devices are prone to different types of attacks from mobile malware.

## 5 Mitigation Strategies

We propose some possible mitigation strategies against attacks from mobile devices. We identify three major reasons for the vulnerabilities: weak anti-malware defense in mobile devices, weak or no encryption and message authentication used in the communication channel, and finally, weak or no defense against DoS attack in medical devices. Table 1 gives an overview of attacks, consequences, and potential mitigation strategies.

**Defense against malware:** If we can take strong defense against malware in the mobile devices, the possibility of attack will be certainly reduced to a great extent. In our threat model, most of the attacks are from malware-infected mobile devices. Without using a malware as a proxy, the adversary needs to be present inside the Bluetooth or WiFi coverage area and also must be paired or connected with the existing Bluetooth or WiFi network. Hence, strong defense against malware will reduce the attack surface and can help us to mitigate the risk of attacks. Unfortunately, due to resource limitation in mobile devices, traditional signature based malware detection softwares are not efficient for these devices. However, some cloud based anti-virus architectures have been proposed [33] to overcome the power restriction. Another approach of detecting mobile malware is monitoring battery usage. When a malware is running in a devices, it must consume some extra battery power. Hence, an anomaly from normal battery usage can be used to detect the presence of a malware [24].

**Low powered strong encryption & authentication:** The two recent incidents of medical device hacking occurred due to an unencrypted communication channel [10, 27]. Manufacturers often skip encryption to minimize power consumption and cost. As many medical devices are battery powered, we need to focus on finding low power consuming strong encryption and authentication schemes. For implantable devices, we need more power efficient strategy because it is not possible to change the battery of these devices frequently. Both Bluetooth and WiFi have their own encryption scheme. However, researchers accomplished several successful attacks on the WEP encryption scheme of WiFi [4, 11, 39]. WPA2 is still considered as secured encryption scheme for WiFi channel, but not all the medical devices are using this encryption due to the power constraint. All the pairing and authentication process in Bluetooth is based on customized SAFER+ block cipher [26] that uses 128 bit key for current Bluetooth standard. However, the heart of the encryption is the PIN number, that is entered by the pairing devices, and it is possible to crack this PIN [35].

**Defense against DoS:** Denial of Service (DoS) and Resource Depletion (RD) attack can be very crucial in health care system. There exists some solutions against DoS attack in IEEE 802.11 network protocol [9, 29], but it is still unexplored whether these solutions are feasible for low-power medical devices.

**Anomaly-based intrusion detection:** For all the attacks that we mentioned here, there must be some unexpected communication between mobile and medical devices. Such anomalous network usage can be used by a intrusion detection module in the medical devices. Again, power restriction will be an obstacle to run this module continuously. We can choose a certain interval or random interval to awaken it. For some devices, such as, blood pressure monitor, we can let the threat detection module run only when the user chooses to do so. Before reading data from display devices, physicians can run this module to make sure whether the data is coming from the actual sensor device or a malware.

| Threat | Consequence | Mitigation Strategy |
| --- | --- | --- |
| Sniffing confidential data from the WiFi or Bluetooth communication channel | Breach of confidential medical record, patient location, and health condition. | Strong low power consuming encryption scheme and Network anomaly based intrusion detection |
| Sending malicious command from mobile devices to medical devices. | A lethal dose to a insulin pump or a stop command to a pacemaker, killing the patient. | Low power strong authentication scheme. |
| Sending misleading information from the mobile devices to the display. | This will misguide the doctor to take appropriate decision. | Low power strong authentication scheme and Network anomaly based intrusion detection. |
| Launching a DoS attack from mobile devices making it unavailable. | Unavailability of critical medical devices can be fatal. | Defense against DoS attack which will be suitable for low power medical devices. |
| Battery exhaustion attack by keeping the communication channel busy. | Can kill patients who use pacemaker or defibrillators. | Network anomaly based intrusion detection. |

Table 1: Overview of threats from mobile device, consequences, and mitigation strategies

## 6 Related Work

Many researchers have explored the vulnerability of wireless-enabled medical devices. Halperin et al. presented a software radio attack on pacemakers and implantable cardiac defibrillators (ICD) [20]. The ICD was designed to communicate in 175 kHz frequency with an external application. They reverse engineered the ICD's protocols to launch attacks on integrity and confidentiality of data. They showed that the communication channel was unencrypted, and it is possible to change the operation of the ICD. They also proved that a battery-powered ICD can be kept busy continuously by an unauthenticated device, which can eventually drain out the battery power.

Radcliffe successfully hacked his own insulin pump [27]. He showed that untraceable attacks against wireless insulin pumps, pacemakers and ICDs are possible from half a mile away. As the transmissions were not encrypted, he was able to decipher the messages. He successfully reverse engineered the application protocol. At first, he successfully jammed the communication channel. Later, he gained full control over the insulin pump and glucose meter. Researchers at McAfee Inc. also successfully compromised an insulin pump [10], and were able to find vulnerable insulin pumps and send lethal doses from a distance of 300 feet.

These attacks were done by reverse engineering the radio channel, without any direct knowledge of the proprietary communication protocol. New generation medical devices communicate over Bluetooth or WiFi channels. The protocol standard of these technologies is open to everyone. Also, Denial of Service attacks or packet sniffing do not require any reverse engineering, which consequently led the security researcher to be more aware of the attack through these popular channels.

Paul et al. identified some security breaches of wireless insulin pump system and proposed mitigation strategies against the threats [32]. They were also able to send command from an unauthorized program [31]. Sorber et al. examined the security of mobile health (mHealth) systems in which personal mobile devices serve as a gateway between the EMR management system and medical sensor devices. They proposed an architecture, *Amulet*, that will ensure privacy and security of mHealth system [37]. Arney et al. described some active and passive attack model on Biomedical devices [5] over wireless channel. They focused on patient's physical security, clinical data security and privacy, and medical device security. They also identified some key challenges for which the security of the medical devices is still a nightmare. Goodman et al. examined the possibility of homicide and extortion attacks by hacking implantable medical devices [17]. He also pointed out the difficulty of investigating this type of homicide cases.

We leverage this body of existing research in exploring the threats to medical devices. The threats from infected mobile devices to medical devices, the core focus of this paper, has not been explored fully by researchers. The closest research related to our work is done by Sorber et al., who proposed a smart card based software security scheme for mobile applications [38]. While their approach allows creating trustworthy applications on mobile devices, they do not consider what will happen if the malware itself is capable of communicating with sensor devices. However, we have already pointed out the strength of the smart devices as a tool for hacking, such as hacking a car's ECU by android phone through the Bluetooth channel [12]. In this paper, we take the first step in exploring the threats from mobile malware to medical devices. Though no real attack has not been yet recorded, we must explore this new threat model to take some early defenses against real attacker.

## 7 Conclusion and Future Work

As mobile devices become smarter, the threats from mobile malware to surrounding infrastructure also increases. WiFi and Bluetooth interfaces enable seamless communication between medical devices and other parts of the healthcare infrastructure. However, this capability has increased the attack surface of medical devices and opens the possibility of novel attacks. We argue that threats on medical devices from mobile devices, especially malware based attacks, are still unexplored, though mobile malware is exploding.

In this paper, we explored the threat of attacks from mobile devices to medical devices. We posit that the intrinsic nature of mobile computing makes the problem difficult: mobile devices are likely to be inside the security perimeter of the healthcare facility's network, be close to the patient, and can be used to launch localized attacks. We also opine that the resource limitation in both medical and mobile devices is one of the main obstacles to defend against these attacks. To protect vital medical devices from attacks, we need to explore the problem domain deeply and develop effective and low-cost solutions. In future research, our goal is to build a proof-of-concept application on Android platform which can be used to evaluate a medical network for vulnerabilities. We also plan to build and test our network anomaly based intrusion detection scheme to detect attacks on medical devices.